\documentclass[submission,copyright,creativecommons]{eptcs}
\usepackage{mathptmx}
\usepackage{epsfig}
\usepackage{comment}
\title{A Framework for Analysing Driver Interactions with Semi-Autonomous Vehicles}
\def\titlerunning{Analysing Driver Interactions}   
\def\authorrunning{Shaikh, S.A.  and Krishnan, P.}
\author{Siraj Ahmed Shaikh
\institute{
Digital Security and Forensics (SaFe) Research Group,\\
Department of Computing, Faculty of Engineering and Computing,\\
Coventry University, Coventry, UK\\
\email{s.shaikh@coventry.ac.uk}
}
\and
Padmanabhan Krishnan
\institute{
Centre for Software Assurance,\\
Bond University, Gold Coast, QLD 4229, Australia\\
\email{pkrishna@bond.edu.au} 
}
}

\begin{document}

\maketitle              

\begin{abstract}
Semi-autonomous vehicles are increasingly serving critical functions in various settings from mining to logistics to defence. A key characteristic of such systems is the presence of the human (drivers) in the control loop. To ensure safety, both the driver needs to be aware of the autonomous aspects of the vehicle and the automated features of the vehicle built to enable safer control. In this paper we propose a framework to combine empirical models describing human behaviour with the environment and system models. We then analyse, via model checking, interaction between the models for desired safety properties. The aim is to analyse the design for safe vehicle-driver interaction. We demonstrate the applicability of our approach using a case study involving semi-autonomous vehicles where the driver fatigue are factors critical to a safe journey.
\end{abstract}

\section{Introduction}

Human failure is often a cause of accidents. Increasing the level of automation while useful in many cases, does not necessarily reduce the number of human failure related accidents. Standards such as ISO 26262 describe functional safety but do not explicitly describe the role of the human user. Most of the research related to ISO 26262 focusses only on the reliability of the electronics and the human user is often ignored. Reliable electronics is not sufficient to guarantee safety.

For such automation to be successful the human user must be aware of the automation and react to it appropriately. In some cases it is not possible to fully automate the behaviour and the system has to rely on humans exhibiting the right behaviour. Examples of such systems include Unmanned Aerial Vehicle (UAV) guidance \cite{Che10}, health care especially patient safety \cite{Car11} and computer security \cite{Cra08}. Thus it is important to understand the role of humans in such systems and analyse the ``human in the loop'' behaviour \cite{Cra08}.

Recent research effort in this direction has seen the choice of various formal methods to model and verify correctness with a view to also modelling human behaviour as it interacts with the system interface. This is largely achieved for isolated behaviours and cognitive errors including mode confusion~\cite{Rus02}, post-completion errors~\cite{CRB07}, error recovery~\cite{SKC07} and automatic behaviour~\cite{Cer11}.

The modelling and analysis of such systems however is non-trivial.
For instance, to analyse driver interactions with a road vehicle requires
knowledge of the vehicle (to model for vehicle dynamics),
the environment (such as road conditions, weather, terrain and traffic flow)
and a model of the human factors that affect the interaction with the vehicle
such as levels of stress, expertise, attention and fatigue. 

In this paper we present a modelling and analysis framework to model human
behaviour and analyse the interaction to determine if any safety conditions are
violated. We model a semi-autonomous system where the driver is part of the
control system. Our approach allows us to incorporate human factors in the
controllability analysis of the system. Our focus is on the interaction between
the user and the system; we do not necessarily focus on the reliability
of the system.

\subsection{Rest of this paper}

The rest of this paper is organised as follows. Section~\ref{motivate:sec} provides motivation behind this work with a view to limitations of some existing work in this area. Section~\ref{approach:sec} presents our approach and describes how it serves to combine the individual strands of effort from Section~\ref{motivate:sec} to provide for a basis for further analysis. Section~\ref{formal-struct-approach:sec} is an attempt to describe our approach in brief formal terms. 

 Section~\ref{casestudy:sec} describes a case study that we use to demonstrate our contribution. The problem of driver fatigue is well understood. Recent developments in driver assistance and adaptive systems means that the safety risks that emerge as a result of interaction between drivers and increasingly semi-autonomous vehicles is yet to be fully explored. System, behaviour and environment models are described in Sections~\ref{system-model-casestudy:sec},~\ref{behaviour-model-casestudy:sec} and~\ref{environment-model-casestudy:sec} respectively. The verification process is described in Section~\ref{verification-casestudy:sec}.

Section~\ref{results:sec} presents the results of the application of our approach to this problem, with notable results discussed in Sections~\ref{sec:lower_speed} and~\ref{sec:man_override}. Section~\ref{conclusion:sec} concludes the paper with a brief comment on the contribution made by this paper.

\section{Motivation}\label{motivate:sec}
 
There are potential economic, health and safety benefits of semi-autonomous vehicles in various industrial applications (e.g., mining \cite{MUP99,FiS12}). Although the level of automation in mining is more advanced than many other domains, human oversight and control is still necessary given various factors such as legacy equipment, interoperability of hardware, and the ability to handle unforeseen circumstances. It is essential to use virtual engineering environments to model the vehicle and environment which can then be used to train drivers \cite{MUP99}.

In addition to the known challenges, such as mode error where the driver cannot recall what state the system is in, there are particular challenges posed by semi-autonomous vehicles that merit attention including 
\begin{itemize}
 \item[-] \textit{handover} between manual and automated control during a task~\cite{Lar10,VHA08}, which is critical as the driver needs to be able to judge when to reclaim control or otherwise, 
 \item[-] \textit{inadequate feedback} from the vehicle to the driver~\cite{Nor07}, with the consequence that the system fails on drivers' expectations during a task and ultimately maximum benefit of the technology is not derived, and
 \item[-] a fundamental change of task for the driver as it changes from monitoring the  \textit{situation} to monitoring the \textit{situation and automation}~\cite{Lar12}.
\end{itemize}

Most of the work done so far in this area has addressed such challenges in isolation and at a high abstract level~\cite{Rus02,CRB07,SKC07,Cer11}, has studied vehicle sensor data~\cite{VHA08}, driver feedback~\cite{Lar12,Mil97,ThB03,OrR07,VSG11,GSO11} (in a real or simulated environment) or performed physiological assessments~\cite{She09,SMC11}. The latter two strands of work entirely focusses on driver perception and experience; moreover borrowing from separate traditions of cognitive and physiological science. 

Another approach to understand the interaction between these various systems is to conduct empirical studies~\cite{CBM12}. It is not possible to conduct an empirical study that includes all these parameters. So most empirical psychological and ergonomic studies focus on a few parameters~\cite{Che10,BMP11,RMS08}. It is hard to visualise a holistic model from such results. As full system interactive behaviour depends on a number of parameters, it is important to have an analysis method that allows specification of all relevant parameters.

Hence the need to allow for more sophisticated models of driver behaviour and vehicle dynamics, on the one hand, and derive such models from multiple disciplines, on the other, to capture the true nature of the problem. The recognition that effective analysis could only be achieved as such is the main motivation behind our work. 

More recent work has acknowledged this. Oppenheim and Shinar \cite{OpS11} develop an approach to model these various aspects that can enable such analyses.  They identify a number of parameters for each model that are important, which is addressed later in Section~\ref{results:sec}. 

\section{Approach}\label{approach:sec}


We adapt the standard discrete-event simulation \cite{RaW89} where we have inputs and outputs and the state of the computation.
We generalise the separation between plant and controller, and augment the controller with a human operator allowing us to model systems that explicitly have human in the loop \cite{Cra08}. This allows us to analyse the interaction between the human operator and the control system to achieve safe behaviours. We use model-checking (and reachability analysis \cite{AlD11}) to determine the safety of the entire  system.

We propose a \textit{system model} to represent the features and behaviours of the system that we deal with. The importance of this model is argued by Bass et al.  \cite{BFG11} as divergence between the state as described by the system model and the user's mental model is often the cause of unsafe behaviour. 

This paper has a focus on semi-autonomous vehicles, and our representation allows for specifying relevant driver control (acceleration and braking) and engine control unit features (odometer and service meters) and chassis control data (vehicle handling and steering, and braking and stability sensors). Aspects of autonomy (adaptive cruise control, lane discipline and navigation) can be factored in as part of the system model. 

A \textit{behaviour model} is then used to demonstrate user actions driven by cognitive and emotional stimuli. The behaviour modelled is an abstraction of the user's mental model and associated actions relevant to the interaction with the system. Traditionally such models have been derived from cognitive science~\cite{Rus02,CRB07,SKC07,Cer11}. However it is increasingly feasible to look to human physiology to sense for driver perception, stress and comfort given advances in sensors \cite{She09}. Recent work carried out quantitative physiological assessment of human stress in response to vehicle interface design (ranging from touch display, voice to multimodal control) \cite{SMC11}. Such an approach provides for an objective assessment of the human condition and further possibility of system adaptation for refined interaction.

Ultimately, a model of behaviour could then be drawn from both branches of science to inform the analysis. Our approach permits the integration of different models. We discuss this issue in our case study section. It is important to note that we do not validate
the behaviour model. We only check if the joint behaviour of the system -- the control system and the human operator -- is safe. One can view the behavioural model as documenting the assumptions we make about the human user.

We also propose an \textit{environment model} to account for operational factors external to the system. Such factors are strictly beyond the control of the system or the user, and remain unaffected by any interaction that results. This provides for a clear separation to study the potential impact on the driver and the system as they interact with the environment individually or while interaction. 

The various factors influencing each of the models are represented as a set of parameters. Some of these parameters would be derived from the cognitive or physiological model underpinning the behaviour model while others would be directly measured.  Assuming there are no circular dependencies, based on the inputs and the current state of the system the outputs and the new state of the system is computed.

\begin{figure}[htb]
\centerline{\epsfxsize = 12cm \epsfbox{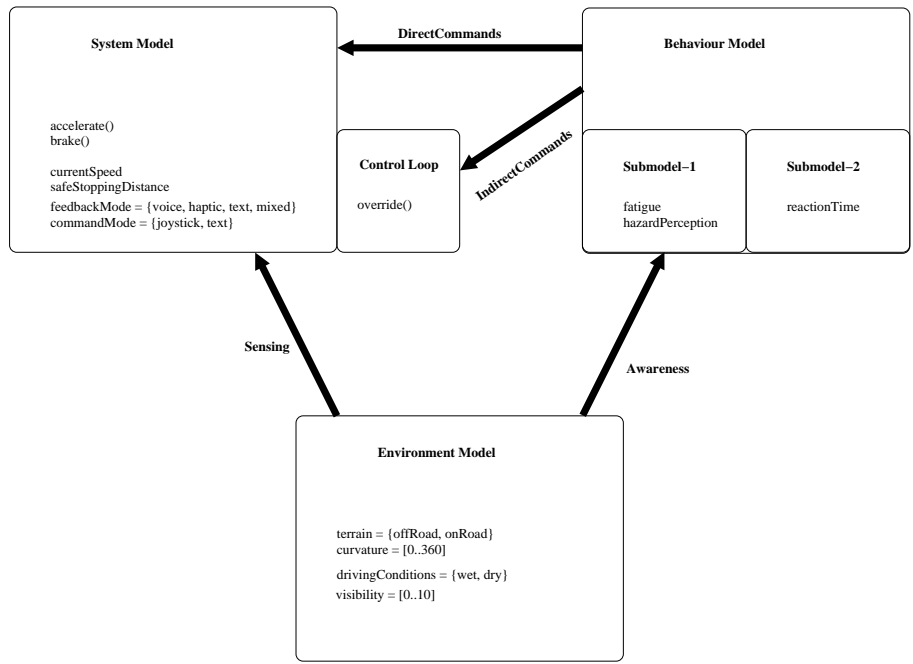}}
\caption{Schematic Representation of Approach.}
\label{schemeModel:fig}
\end{figure}

A schematic description of our approach is given in Figure~\ref{schemeModel:fig} and the behaviour of the entire system is expressed using the control loop shown in Figure~\ref{cloop:fig}. In our approach we assume that within one iteration the environment can affect the system and the environment and system can affect the user's behaviour.

\begin{figure}[htb]
\begin{verbatim}
loop {
   calculate values from the environmental model
   calculate values from the system model
   calculate values from the behavioural model
   calculate the outputs
   update the state of the system 
   update the state of the user  
   assert(properties)
   generate the_outputs
}
\end{verbatim}
\caption{Control Loop.}
\label{cloop:fig}
\end{figure}

The entire system is developed in C and we use \textit{CBMC (Bounded Model Checker)} \cite{CKL04} to validate the safety properties. Given that it is a bounded model-checker we either have to specify the number of iterations that need to be explored or have to set up the system such that it can calculate the bound. Thus when CBMC indicates that a system is safe, it is safe only within the bounds specified for the verification. However, if it finds an error (that is, a counterexample that violates the assertion) we can be sure that the system is unsafe.

Our approach allows us to explore various scenarios where safe interaction between the driver and vehicle is critical. Given behaviour, vehicle and environment models, overall system states that lead to safety violations would be explored. Specific scenarios can be generated if the functions that represent the calculation of the model values return a unique value. We also use the CBMC feature that allows a function to return non-deterministic values which effectively allows us to analyse a class of behaviours.

\subsection{Formal Structure}
\label{formal-struct-approach:sec}

In this section we present some of the formal details associated with our
models and the verification process.
While the exact details will depend on the system being studied, there are some
general principles that underly our approach.
We focus on the behavioural and environment models. The system model is
straightforward -- we assume there is a linear ordering of the variables
which is used by the control loop.

The environment model is to used to explore different scenarios.
We can either set the variables in the model to specific values or choose
(non-deterministically) from a collection of possibilities.

The empirical results that we use are represented as tables.   Thus
the basic values in the behaviour model are obtained via a table lookup.
If there are multiple models that determine a particular value, one has to
define a suitable function that combines the values. Choosing this function 
non-deterministically allows exploration of situations where empirical results
are not available.

The environment model consists of variables $E_I, E_O$ and $E_S$, the system model
consists of variables $S_I$, $S_O$ and $S_S$ while the behaviour model
consists of variables $B_I$, $B_O$ and $B_S$. 
We use the subscripts $O$, $I$ and $S$ to denote output, input and state
respectively.

There are two types of assignment statements. The first (of the form $x = e$) evaluates
the expression $e$ and assigns the value to the variable $x$. The second (of the form
$x = nondet(P)$) chooses a value from the set of possibilities ($P$) and assigns to $x$.
As we are using the CBMC model-checker  to verify behaviours this is equivalent
to checking each and every  value from $P$.

The control loop (shown in Figure~\ref{cloop:fig}) for the system consists of the following steps.
\begin{enumerate}
\item $E_I' = f_{EI}(S_O,E_S)$  where $f_{EI}$ is a function that calculates the new input values
for the environment and $E_I'$ denotes the new set of input values.
The user can influence the environment only via the system and hence the outputs
of the behaviour model are not used in this calculation. The update function may involve
non-deterministic choices.
\item $S_I' = f_{SI}(B_O,E_O,S_S)$  and
\item $B_I' = f_{BI}(B_S,E_O,S_O)$  are similar to the first step. The only restriction is that the
system update function is deterministic.

\item The next three steps calculate the outputs of the three models, viz.,  
\begin{description}
\item $E_O' = f_{EO}(E_I',E_S)$,
\item $S_O' = f_{SO}(S_I',S_S)$,
\item $B_O' = f_{BO}(B_I',B_S)$.
\end{description}

\item The next three steps update the local state of the models. They are,
\begin{description}
\item $E_S' = f_{ES}(E_I',E_S)$,
\item $S_S' = f_{SS}(S_I',S_S)$,
\item $B_S' = f_{BS}(B_I',B_S)$.
\end{description}

\item Execute the assertions that encode the safety property.
\end{enumerate}

The various functions shown above essentially capture the three models in our system.

\section{Case Study: Driver Fatigue}\label{casestudy:sec}

Enhancing the driver experience through increasing autonomy has been of interest for well over a decade now. Reduction in drive stress, freeing up limited attentional resources and improving road safety have been the major goals of this effort. However, autonomy brings with it a variety of other challenges that potentially risk road safety \cite{StM96}. This could be due to sensor limitations, system design faults, error inducing design, or inadequate driver training; these certainly are some of the lessons learned from the introduction of autonomy in the aviation domain.

Of interest here is driver fatigue, which results in drowsiness and hypovigilance, particularly after prolonged periods of driving and monotonous roadside experience through motorways \cite{ThB03}. This has been confirmed by a number of studies \cite{Mil97,LDDW98}. This is a major cause of accidents across the world, with around 10-35\% of all road accidents in the USA and several European countries estimated to be fatigue and sleep related~\cite{RoSPA01}. While manual strategies adopted by drivers to cope with this problem are recognised \cite{GSO11}, fatigue serves to be one of the main factors for increasing autonomy in vehicles \cite{OpS11}. 


We view each  vehicle as a safety critical system
where one should avoid accidents as well as complete the mission. 
Even if the vehicle is unoccupied, a human will be involved in its control.
The vehicles are travelling in a convoy and it is important to maintain
safe stopping distance. Otherwise, unexpected environment
events (like explosions) can cause the driver or human controller to accelerate
and run into the vehicle in front of the convoy.
The case study studies the effect of
driver ability and input modes on the desired vehicle
separation  for safety given a specific route.

The example we present here 
focusses on using different actual empirical results
that are incorporated into our behavioural
models. For instance, the parameters for
hazard perception are chosen from \cite{CMC10,Che10}.
Similarly, the parameters for the control mode are chosen
from \cite{BMP11,CMC10}
while the
desired separation and safe stopping distance  are chosen from \cite{CMC10}.
The function that calculates the reaction time from
factors such as fatigue is derived from \cite{GrM11,BMP11,CBM12}.
Other factors such as speed and route are generated (either manually or
non-deterministically) as part of the scenario exploration experiment.

The use of sub-models requires
us to experiment with hypothetical integration to achieve
a single behavioural model.
We have to consider hypothetical integration
due to lack of suitable models that cover all aspects covered
in the sub-models.  This separation
of actual models and hypothetical integration
documents the assumptions under which our analysis is valid. It also points
to areas where more precise cognitive models are required.

\subsection{Structure of the ACC System Model}
\label{system-model-casestudy:sec}

The system model has the key aspects of the ACC mainly the control
system to maintain speed of the various vehicles, the
separation between them and a notion of safe stopping distance to prevent
accidents.
The environment model has the terrain and hazards that are non-deterministically
generated.
The behaviour model we consider is that of fatigue which is
calculated from the duration the driver has been in the vehicle and
the complexity of the terrain.
The driver's cognitive ability is influenced by fatigue and hazard perception.
This
when combined with the mode to issue the commands determines the reaction
time. The current speed of the vehicle and the reaction time of the driver
will determine the safe stopping distance.

Adaptive cruise control (ACC) is one of the mechanisms introduced to provide safe distance control from lead vehicle in front: once engaged, the vehicle operates in a typical cruise controlled fashion with the added feature of sensing the vehicle in front to adapt speed (if it slows down or speeds up) ensuring a minimal safe distance at all times. Figure~\ref{accalgorithm:fig} below shows an algorithm for a simple implementation of ACC. 
\begin{figure}[htb]
\begin{verbatim}
void function ACC(dist_veh_in_front)
{ 
   loop{
        input current_dist_veh_in_front; 
        IF dist_veh_in_front == current_dist_veh_in_front
         CALL Maintain_host_veh_speed;
        END IF
        IF dist_veh_in_front > current_dist_veh_in_front
         CALL Decelerate_host_veh_speed;
        END IF
        IF dist_veh_in_front < current_dist_veh_in_front
         CALL Accelerate_host_veh_speed;
        END IF
   }
}
\end{verbatim}
\caption{Adaptive Cruise Control (ACC) algorithm in pseudocode}
\label{accalgorithm:fig}
\end{figure}
The \verb|ACC| algorithm uses the vehicle's on-board sensors to read in the gap from the vehicle in front. This is preset by the driver and passed onto the algorithm as a parameter. Once it enters the loop, it strives to maintain this distance by continuing to maintain the vehicle speed if the gap to the vehicle in front is the same, decelerating if the distance gets narrower than desired or accelerating if the gap is wider.  

Studies have demonstrated that ACC has the potential of causing delayed driver reaction~\cite{VSG11}, and awkward handover and mode confusion with up to a third of drivers having forgotten at some stage whether ACC was engaged or otherwise~\cite{Lar12}. This has serious road safety risks and raises a question whether the design of such mechanisms would ultimately serve to be detrimental to the intended goal. In addition to the time on task effects, road conditions and terrain also significantly affect driver experience, and contribute to fatigue \cite{OrR07};
difficult terrains require more frequent driver interventions \cite{SAY04} in semi-autonomous vehicles.

The above points to a clear need to analyse the relationship between driver control and vehicles offering semi-autonomous features such as ACC. The case study proposed to demonstrate the framework presented in this paper revolves around how driver behaviour and perception is affected
by fatigue which itself is influenced by factors such as
journey time. The behaviour model is influenced by parameters including over time on task, fatigue and perception of ACC status. The system model is parameterised to represent a vehicle equipped with ACC, ranging over speed, acceleration, distance to vehicle in front, and ACC status.

\subsection{Structure of Behaviour Models}
\label{behaviour-model-casestudy:sec}

The sub-models associated with the behaviour model are now described.
The fatigue is calculated using the variables that are
identified by  Oppenheim and Shinar \cite{OpS11}. 
Although they do not present empirical results they summarise
results from other papers.   These results are not specific to the
military situation. They are more from generic  driving conditions. However,
as  that is the best data available we use it. 
Similarly, the role of terrain is identified as an important factor
\cite{CBM12,RPI08}. The role of terrain is explained only in qualitative
terms and neither present any concrete model of how the terrain actually affects
fatigue.
Therefore, there is no obvious way to create a unified
behaviour model from the empirical results.

Thus we explicitly define an integrator function that will combine the variables
and values indicated in prior work \cite{OpS11,CBM12,RPI08}.
As the behaviour of this function is not available,
we encode various  candidate functions. 
All candidate functions are automatically evaluated in the verification
process. Technically,
this is achieved using non-deterministic choice in CBMC.
 So
if our analysis indicates that safe stopping distance is
always maintained, we can conclude that safety is independent of
the candidate functions.

The reaction
time calculation is from Baber et al. \cite{BMP11}.
The reaction time depends on the mode of communication
between the driver and the vehicle control system as well as the derived
value for the fatigue factor.
Baber et al. \cite{BMP11}  present empirical results
for mode of communication and reaction time. 
Here again we rely on a non-deterministic choice
of integrator functions to combine the model from Baber et al.
\cite{BMP11} and the calculated fatigue value.
For the case study the interaction mode is a factor where research
shows that in certain circumstances the use of speech is better than using text
while the use of joystick or gamepad to control the system is often
better than issuing text based commands.

For instance, the empirical results from Baber et al. \cite{BMP11} are encoded
to calculate the reaction time is shown in
Figure~\ref{encodingHumanFactors:fig}.
They function is  \verb|setReactionTime| which calculates a qualitative
reaction time based on the input mode and then invokes
\verb|changeReactionTime| with the fatigue level to calculate a quantitative
value.

\begin{figure}[htb]
\begin{verbatim}
void setReactionTime(mode iM, fatigueLevel df)
/* mode is the input mode used by the driver and
   fatigueLevel is a indication of the driver's  tiredness
   we use symbolic values rather than concrete values for the reactionTime
   
*/
{
  switch (iM) {
    case GamePad: reactionTime = okay;
    case Speech: reactionTime = slow;
    case MultiModal: reactionTime = fast;
  };
  changeReactionTime(df);
}

void changeReactionTime(fatigueLevel df)
/* this function alters the reaction time based on the
   level of tiredness. These factors can either be chosen from
   empirical results  or can be  set non-deterministically
*/
{
   switch (df) {
     case Exhausted : reactionTime = reactionTime * eFactor;
     case Tired     : reactionTime = reactionTime * tFactor;
     case Normal    : reactionTime = reactionTime * nFactor;
    };
}
\end{verbatim}
\caption{Encoding of Human Factors.}
\label{encodingHumanFactors:fig}
\end{figure}

In general we represent the parameters in the various models as
global variables and represent the actual calculation
of the the values as functions. For instance,
\verb|timeDriven| and \verb|terrain| are global integer values.
As there are discrete levels of fatigue
we use enums to represent the possible values
the global variable \verb|driverFatigue| can take.
The function \verb|setDriverFatigue| assigns a value to the variable
\verb|driverFatigue| based on the behaviour model.

The hazard perception is based directly on Chen \cite{Che10}. As we are using
only a single model there is no need to define an integrator for hazard
perception.
This ability includes the size of the hazard.
In general larger the hazard the easier it is to perceive.
We use the variable \verb|hazardPerception| to denote the
user's ability to perceive hazards. This can be derived
from \verb|driverFatigue|. It is also possible
for  one to use CBMC's non-determinism to generate a value.
An example is  the following statement
\begin{quote}
\verb|hazardPerception = hpFunction(nondet_int());|
\end{quote}
where the function
\verb|hpFunction| converts an integer
to a suitable hazard perception value. 
The function \verb|hpFunction| is a part of $f_{BI}$ in the formal model.

\begin{figure}[htb]
\begin{verbatim}
v1 = model1(p1); /* p1 represents the parameters used by
                     the first model */
v2 = model2(p2); /* p2 represents the parameters used by
                     the second model */
opChoice = nondet_uint() % choices;
return choiceFunction(opChoice,v1,v2);
\end{verbatim}
\caption{Non-Deterministic Integrator Function}
\label{nondetI:fig}
\end{figure}

The structure of a non-deterministic integrator function is
shown in Figure~\ref{nondetI:fig}.
The variable
\verb|opChoice| represents the function to combine the values from the
different models and its value is calculated non-deterministically. 
The actual integration is performed by the
function \verb|choiceFunction| which applies the chosen function to
the values from the sub-models to return the actual value for the
joint model and
represents a specific instance of $f_{BS}$ (i.e., the function that calculates the state associated with the user's behaviour).

\subsection{Structure of Environmental Models} 
\label{environment-model-casestudy:sec}

The environment model allows for variability in scenario to help analysis,
including road conditions, terrain or obstacles
that result in manual driver interventions) and journey time.
The travel scenario is expressed as a sequence (i.e., an array)
of  route points where each
route point has the relevant information (and represented as a struct).
A simple example of the structure of the route is shown in
Figure~\ref{route:fig}.

\begin{figure}[htb]
\begin{verbatim}
struct routePoints{
int obstacle; /* represents difficulty to overcome */
int distance; /* time to travel is calculated using the speed */
int terrain; /* represents offRoad, onRoad */
int curvature;
};

struct routePoints route[5];  /* there are 5 sections to travel */
\end{verbatim}
\caption{Example of Route.}
\label{route:fig}
\end{figure}


The cause and effect relations, informed by the literature above, are explored to analyse for relative safety thresholds that arise from the interaction of the three models. Potential safety violations in terms of dangerous levels of fatigue and increased likelihood of mode confusion are studied.

\subsection{Verification Process}
\label{verification-casestudy:sec}

In the context of our framework (shown in Figure~\ref{schemeModel:fig})
the ACC is part of the control loop associated with the system model.
The user's commands are, normally, filtered via the control loop except
when the user can override the ACC and issue commands directly to the system.
This type of behaviour is typically seen when the user is fatigued or 
encounters a hazardous situation.

The behaviour of the system (i.e., the vehicle and the human operator)
was analysed for under various conditions (e.g., route choices, travelling
speed).
We were able to show that a small separation between vehicles
was safe if the driver
was not too tired and the interaction mode involved game pads.
We were also able to show that even a large separation was unsafe
if the driver was tired and the interaction mode involved only speech.
The safety conditions are written as assertions such as 

\begin{verbatim}
assert(isOkay(driverFatigue,hazardPerception,safeStoppingDistance));
\end{verbatim}
where
\verb|isOkay| determines if the current separation is
safe given the parameters from the behaviour  model.
These are specific to the application being analysed.

In the function \verb|main| we invoke the function
associated with the control loop (shown in Figure~\ref{cloop:fig})
 with the various
scenarios.
By varying the variables that are assigned non-deterministic
values, one can explore the safety of different scenarios within a given
model.
We also use CBMC to calculate values where failures occur.
For instance, the following program fragment
determines  if there are any
values of \verb|driverFatigue| that can lead to  accidents.

\begin{verbatim}
int driverFatigue = nondet_uint()%numFatigueValues;
assert(isOkay(driverFatigue,hazardPerception,safeStoppingDistance));
\end{verbatim}

CBMC will calculate a value for \verb|driverFatigue| 
 that will violate the assertion. If CBMC cannot find such a violation,
we are sure that the composite behaviour represented by the models
is safe.

\section{Results of Case Study}\label{results:sec}

We evaluate our approach to show how effective it is in capturing the interaction between the various inputs provided to the sub-models. The resulting analysis should allow us to check whether the composite system demonstrates safe behaviours. We consider three scenarios for analysis, which are essentially an exploration into the relationship between fatigue and the various parameters that can be influenced. Evidence suggests that prolonged periods of driving contribute to an increase in fatigue \cite{ThB03,Mil97,LDDW98} and mechanisms such as ACC available to address such problems need to be evaluated for their effectiveness and safety. 

We recall a few select inputs and variables of the system of interest to us here. $Route$ denotes the distance travelled by the vehicle over a given journey, and $speed$ is the speed of the vehicle travels at. The $control~mode$ is used to signal whether the vehicle is driven manually or with ACC enabled. The $safe~stopping~distance$ is the estimated safe distance between the host vehicle and the vehicle in front under the conditions of fatigue. This is related to $desired~separation$ which is understood to be the minimum safe distance (ranging over given vehicle speed). We represent the driver state using a combination of variables including $fatigue$ to show the level of driver tiredness, $hazard~perception$ to show the driver's ability to perceive hazards encountered enroute, and $reaction~time$ as a measure of the time taken for the driver to respond. The variables are defined in relative terms for symbolic representation.

We manipulate the above parameters in different analyses by configuring them as constants where we intend them to be fixed for the purpose of the scenario. To range over values, we use non-deterministic assignment. In cases where values are determined by the respective models, we leave the variables assigned accordingly, either as a result of a fixed model (where assignments are straightforward) or as a result of sub-models combined using a variety of functions chosen non-deterministically.

Note that our implementation of the ACC is adapted to be sensitive to driver fatigue over the course of a journey. Implicit here is that the speed of the vehicle is allowed to be adjusted and the desired separation distance is also left for calculation accordingly. The rule is that as the journeys progresses, the desired separation is extended to account for the increase in fatigue, which in turn contributes to weakened hazard perception. 

In all cases we analyse for two properties. First, that the desired separation is always maintained for the given control mode and level of fatigue. Secondly, that the use of ACC does not result in an actual increase of driver fatigue. 

\subsection{Lowered speed and increased fatigue}
\label{sec:lower_speed}

Our first scenario deals with an unexpected side-effect of the ACC operation whereby the ability to adapt vehicle speed results in an increase in journey time, and hence fatigue.  

We configure the system for a non-deterministic route, and leave the level of driver fatigue to be derived by the system output. We want to check whether the driver fatigue goes past a threshold. The control mode remains enabled for ACC, and the safe stopping and desired separation distances also remain fixed. We let the speed of the vehicle adjusted as per the operation of the ACC. 

Our analysis reveals that the scenario fails to satisfy the second property. To maintain the desired separation distance the vehicle speed is reduced by the duration that increases. This leads to increased driver fatigue. At the next point on the route the safe separation may need to be increased owing to increased fatigue. This in turn further reduces the speed. After a few iterations the vehicle speed is slowed to such a level that fatigue due to time on road becomes unacceptable.

\subsection{Manual override and variable speed}
\label{sec:man_override}

An alternative scenario is where the driver's ability to override ACC at any stage of the journey is acknowledged. This is essentially to model for cases where driver actions may have undesired consequences. 

We consider a fixed route, and the safe stopping and desired separation distances also remain fixed. We allow for the control mode to be non-deterministic and have no control over the choice operated. Driver fatigue is then a calculation based on various inputs from the behaviour and system models. Speed is also dependent on the choice of control mode and ultimately the driver, and so are the rest of the variables. 

Our analysis reveals that the first property is violated. Counter examples are due to a case where the driver is able to manually override ACC and increase vehicle speed, which results in unsafe distance from the vehicle in front.
A different possibility is where the driver switches over to manual mode and ultimately reaches an unsafe state (due to fatigue for example). In one sense such a possibility is difficult to avoid, unless speed or proximity alerts are
modelled. 

\subsection{The ideal scenario}

We consider a final scenario where we control the parameters for a best case scenario: the route and control mode (ACC) are both fixed. Driver fatigue is calculated as influenced by a combination of system, behaviour and environmental models. All other parameters are calculated from the relevant model respectively. 

We are able to confirm that both properties are satisfied. The fixed length of route means that journey time is ultimately limited, even if speed is adapted (slowed) in response to driver fatigue as discussed in Section~\ref{sec:lower_speed}. The fixed control mode helps to avoid any driver-led errors as in Section~\ref{sec:man_override}.

\subsection{Discussion}

For all scenarios, CBMC is able to verify the safety property or calculate a
counter example in less than five seconds on a low end machine
(an Intel U1400 processor running at 1.2GHz with 1.0GB RAM running Linux).
Our typical loop unwinding parameter is 100 which is enough to explore all behaviours. Unfortunately, given the calculations involved CBMC is unable to prove the unwinding assertion.

There are two main limitations of this case study. The first is getting an appropriate system model as such models are largely proprietary. We have constructed this model from various published sources.

The second is the choice of ``sensible'' ways to combine sub-models. While our framework can explore any function, it is not clear what classes of functions are close to reality. We have explored only simple arithmetic operators for the two integrator functions mentioned earlier, namely, the safe stopping distance and fatigue. These operators can be chosen non-deterministically. This is a clear limitation in our case study. Technically the integrator functions need empirical validation but owing to the number of parameters such validation studies are hard to perform.

\section{Conclusion}\label{conclusion:sec}

The main contribution of this work is an approach to
modelling and design of human in the loop systems. The approach takes into
account real systems as well as cognitive models  that are supported
by empirical studies.

By expressing the  models in the C programming language we are able
to encode all empirical  models without compromising on numerical accuracy.
We have had informal discussions with practising engineers from the automotive
domain and the general feed back is that the approach is simple enough to be
used by them. Their models are expressed as code fragments that can be
translated quite easily to C.

We also avoid concurrency related issues using a standard sense-control
cycle approach. This simplification (which reduces the state space that needs
to be explored) when combined with
CBMC enables us to handle larger realistic models.
The only care required is the sequence in which the variables are updated.
Using C is an advantage of our approach over other approaches that use specific modelling languages \cite{CRB07,BFG11}.

This approach supports automatic verification of safety properties as well
as systematic scenario exploration.
The non-deterministic choices of the various functions
in the behaviour sub-models document the assumptions that we
make on the interaction between different aspects. These interactions
are not supported by empirical studies but occur in real systems.

The case study was chosen to demonstrate our approach. 
Further work beyond this early exploration would look to both adopt more mature models of behaviour with respect to semi-autonomous vehicles~\cite{SYo00,SYo05}, and draw parallels to research that has addressed intricate problems in this area from other domains~\cite{MMB01} which have traditionally relied on driver experience and feedback to evaluate such vehicles. 

We are also currently exploring the encoding of more specific models,
for instance, where the adaptive cruise control system is actually available
(and hence we do not need to model it) or automobiles with
in-wheel motors where the dynamics are more complex. 
The aim is to integrate the formal analysis of user interaction
with relevant standards for reliability including the automotive safety
integrity level requirements.

\subsection*{Acknowledgement}

We are grateful to Graham Shelton-Rayner and Damian Harty for useful discussions. Padmanabhan Krishnan was partially supported by a grant from Coventry University, UK.

\bibliographystyle{eptcs}
\bibliography{refs}

\end{document}